\documentclass[aip,graphicx]{revtex4-1}
\usepackage{graphicx}
\usepackage{dcolumn}
\usepackage{bm}

\usepackage[utf8]{inputenc}
\usepackage[T1]{fontenc}
\usepackage{mathptmx}
\usepackage{etoolbox}
\usepackage{multirow}
\usepackage{hyperref}
\usepackage[capitalise]{cleveref}
\usepackage{caption}
\usepackage{subcaption}

\draft 

\begin{document}


\title{Study of the UHV compatibility of selected ferritic stainless steels for application in vacuum systems of future gravitational wave detectors} 



\author{Carlo Scarcia}
\email[email address: ]{carlo.scarcia@cern.ch}
\author{Giuseppe Bregliozzi}
\author{Paolo Chiggiato}
\author{Ivo Wevers}
\affiliation{CERN}

\date{\today}

\begin{abstract}
Next-generation gravitational wave detectors (GWDs) such as the Cosmic Explorer and Einstein Telescope demand extensive ultra-high vacuum systems, making material cost and performance critical considerations. This study investigates the potential of ferritic stainless steel as a cost-effective alternative to the commonly used austenitic stainless steel for UHV components, focusing on the analysis of outgassing rates pre and post-bakeout at 80°C and 150°C for 48 hours.
The tested ferritic stainless steels exhibit significantly lower hydrogen content than standard AISI 304L steel. After bakeout, the hydrogen outgassing rates — measured down to 10$^{-15}$ mbar l s$^{-1}$ cm$^{-2}$ — are three orders of magnitude lower than those of similarly conditioned austenitic stainless steels. These results highlight ferritic stainless steel as a promising, economical, and high-performance candidate for future GWDs vacuum systems.
\end{abstract}


\maketitle 

\section{\label{sec:level1} Introduction}
The direct detection of gravitational waves in 2015 \cite{firstdetection} opened a new way to observe the universe and revolutionized astrophysics, marking a significant advancement in our understanding of the cosmos.
The ultra-sensitive laser interferometry employed in the gravitational wave detectors (GWD) is used to measure the space-time deformation attenuating all noise sources \cite{noises} that might hinder the measurements.

One of the sources of noise originates from the residual gas molecules present along the laser paths, which can extend for several kilometres. To reduce statistical fluctuations in the number of molecules in the volume through which the laser beam travels, it is essential to maintain pressures in the ultra-high vacuum (UHV) range. This helps minimize variations in the refractive index and the phase shift of the photons \cite{mikegas}. The UHV conditions are then designed to ensure that the pressure-related noise along the interferometer's arms is at least an order of magnitude lower than the sum of all the other noise sources (seismic, Newtonian, thermal, etc.).

The current generation of GWDs (such as LIGO \cite{aLIGO}, Virgo \cite{aVirgo}, and KAGRA \cite{KAGRA}) has marked the 200$^{th}$ detection of the last operation run \cite{200th}, and they continue to provide valuable insights through ongoing enhancements. However, their capabilities remain limited as in a couple of decades, they will reach their ultimate sensitivity.

To broaden the scope of potential discoveries, a third generation (3G) of gravitational wave detectors (GWDs) is being proposed, in particular the Cosmic Explorer (CE) in the United States and the Einstein Telescope (ET) in Europe.
The CE design includes two L-shaped interferometers hosted in distinct locations with arm lengths of 40 km. CE would require approximately 160 km of vacuum tubing with an internal diameter of 1.2 meters \cite{CE}.
ET, on the other hand, in its original configuration, will consist of six interferometers arranged in an equilateral triangular underground tunnel with sides measuring 10 km. This layout will require about 120 km of UHV piping with an internal diameter of 1 meter \cite{ET}.

The considerable size of the vacuum piping system could significantly impact the overall experiment cost, prompting the need for revised designs, fabrication methods, and materials relative to the current generation of GWDs\cite{ETcost}. Therefore, it is essential to select materials and fabrication processes that not only ensure ultrahigh vacuum performance but also optimise costs with respect to the austenitic stainless steel systems currently used in gravitational wave observatories \cite{virgosteel}.

In a previous work, the authors explored the UHV compatibility of mild steels as a potential alternative to austenitic stainless steel for the fabrication of UHV chambers \cite{mine}. The findings revealed that the tested mild steels exhibited significantly lower outgassing rates, outperforming by several orders of magnitude standard UHV-compatible austenitic stainless steels that are not vacuum or air fired. However, despite its lower raw material cost and demonstrated UHV compatibility, concerns remained regarding its susceptibility to internal surface changes and limited corrosion resistance. Addressing these issues may require additional treatments or coatings, which could diminish the mild steel's overall cost advantage.

Another alternative candidate to substitute the austenitic stainless steel currently used for GWD could be ferritic stainless steel. Characterised by a fully ferritic microstructure at room temperature, these stainless steels are not commonly used in accelerators and surface science equipment due to their soft magnetic properties \cite{magnetic} and rapid ductile-to-brittle transition below room temperature, which hinders their use in cryogenic applications as well \cite{cryo}. 
Indeed, the few studies of a UHV application of ferritic stainless steel involved shielding a particle beam from an external magnetic field. The authors of the study report a hydrogen content of 0.04 ppm weight in AISI 430, a ferritic stainless steel grade used to produce ISO flanges \cite{SUS430kamiya}. The content, extracted via Temperature Programmed Desorption (TPD) measurements, was found to be 12 times lower than that in AISI 316L. In a related study, the water outgassing rate after 100 hours of pumping for AISI 430 was found to be similar to that of AISI 304 \cite{SUS430kato}. Unfortunately, these studies did not directly measure the hydrogen outgassing rate. If these rates are as low as those reported for mild steels, ferritic stainless steels will offer the same benefits as mild steel while also addressing the corrosion problems associated with low-carbon structural steels.

This study aims to evaluate ferritic stainless steel as a potential material for 3G GWDs vacuum tubes. We measured the outgassing rates of well-defined ferritic stainless steel grades for the typical gas species found in UHV environments, including H$_{2}$, H$_{2}$O, CH$_{4}$, CO, and CO$_{2}$. To ensure cost-effectiveness, we utilized only readily available off-the-shelf products and assessed the impact of low-temperature bakeouts on outgassing rates. TPD measurements were employed to estimate the H$_{2}$ content in the analyzed samples.
\section{\label{sec:level2}EXPERIMENTAL APPARATUS AND METHODS}
A detailed description of the experimental apparatus and measurement steps can be found elsewhere\cite{mine}. Below, we report the relevant information and changes to the systems used for this study.
\subsection{THROUGHPUT METHOD}
The outgassing rates of unbaked samples were measured using the throughput method \cite{redhead,chiggiato}.
Pumping speed was provided through an orifice with a diameter of 0.8 cm, which results in conductance for water vapour of C$_{H_2O}$ = 7.4 ls$^{-1}$.
The speciﬁc outgassing rate was monitored for approximately 100 h of pumping, with the room temperature stabilised at 21±2°C. The background measurements were repeated every time a new steel grade was tested.

\subsection{\label{sec:accu}COUPLED ACCUMULATION-THROUGHPUT METHOD}
The coupled accumulation-throughput method \cite{chiggiato} was employed to measure the post-bakeout outgassing rates of the samples.
To minimise gas sources other than the analysed samples, all components that constitute the measuring system and sample holder were vacuum fired for 2 h at 950°C \cite{chiggiato} prior to installation. The sample's geometrical surface area was maximized to increase system sensitivity.
The sample holder was baked at 80°C and 150°C for 48 h, while the rest of the system was baked at temperatures ranging from 200°C to 350°C. The accumulation measurements started when the samples were at room temperature (21±2°C), 24 h after the end of the bakeout cycle. 
\subsection{TEMPERATURE PROGRAMMED DESORPTION}
Temperature Programmed Desorption (TPD) measurements were conducted to determine the diffusible hydrogen content. 
The TPD analyses were performed using a commercial TPD workstation\cite{TPDhiden}.
The samples, each having a surface area of 2 cm$^2$ and thickness varying from 0.15 to 0.3 cm, underwent heating from 25°C to 940°C at a ramp rate of 5°C/min. To ensure measurement reproducibility, the background was re-measured after every ten samples. The accuracy of quantitative measurements was verified through regular in-situ calibration of the RGA. The hydrogen concentration calculation relied on sample weight, measured with a weight scale with a sensitivity of ±0.1 mg.
\clearpage
\section{MATERIAL SELECTION AND SAMPLE PREPARATION}
AISI 441 and AISI 444 off-the-shelf sheets were acquired from the market and compared with AISI 304L stainless steel sheets available at CERN for UHV applications. The characteristics of the selected steels are listed in Tab. \ref{tab:aisigrades}. 
The samples for the analysis were dry-cut to the required dimensions using a bench shear. 
Both the ferritic grades and the AISI 304L samples were cleaned following the CERN UHV standard procedure\cite{304Lclean} that implies the use of a detergent bath, rinsing with de-mineralized water, and drying in an air furnace at 60°C for 10-60 min.
\begin{table}[!t]
\centering
\caption{Chemical composition (wt.\%) of the selected stainless steels with the corresponding manufacturing process (MP), heat treatment (HT), surface finish (SF) and shape. AISI 304L chemical composition values are to be intended as the maximum content allowed\cite{304LEDMS}. CR: Cold Rolled, HR: Hot Rolled, RA: Recrystallization Annealed, SA: Solution Annealed.}
\label{tab:aisigrades}
\begin{tabular}{ccccc}
\hline \hline
 & AISI 441 & AISI 444 & AISI 304L\\
\hline
MP & HR - CR & CR  &CR \\
\hline
HT & RA & RA & SA \\ 
\hline
SF & 2B - 2D & 2D & 2D \\
\hline
Shape & Sheet & Sheet & Sheet\\
\hline
C	&0.015	&0.011	&0.03 \\
Mn	&0.385	&0.3	&2.0\\
Si	&0.584	&0.38	&1.0\\
S	&0.001	&0.0014	&-\\
P	&0.0029	&0.0029	&0.03\\
N	&0.014	&0.016	&0.02\\
Cr	&17.57	&18.89	&17-20\\
Ni	&0.242  &-		&10-12.5\\
Mo	&-	    &1.892	&-\\
Ti	&0.16	&0.006	&-\\
Nb	&0.4	&0.566	&-\\
Fe	&Remainder	&Reminder	&Remainder\\
\hline \hline
\end{tabular}
\end{table}
\clearpage
\section{Results and discussion}
\subsection{Water vapour outgassing rate}
\begin{figure}[!h]
\includegraphics{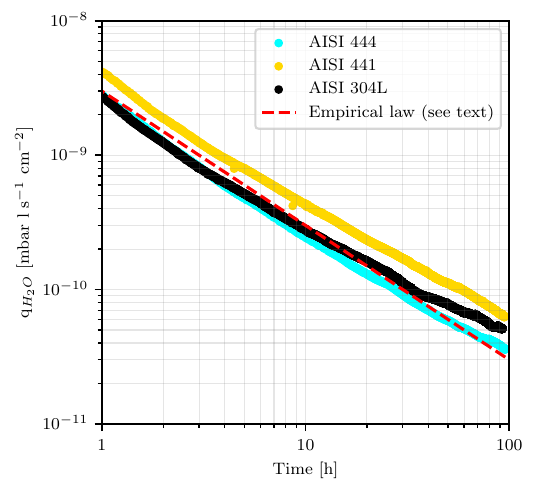}%
\caption{\label{fig:pumpdown}Pumpdown curves at 21±2°C; the specific outgassing rate is plotted as a function of the pumping time. The background value is subtracted. The empirical law, calculated as reported in the text, is generally applied at CERN for UHV-compatible austenitic stainless steels.} %
\end{figure}

The measured pumpdown curves, representing the water vapour specific outgassing rate as a function of pumping time, are shown in \cref{fig:pumpdown}. The specific outgassing rates recorded after 10 h of pumping are summarized in Table \ref{tab:pumpdown}.
\\
As shown in Figure \ref{fig:pumpdown}, the pumpdown curves of the ferritic steel grades are reasonably fitted by the inverse laws of pumping time, as compared with the empirical law (dashed line)
$q_{H_2O}$ = $\frac{3\times10^{-9}}{t[h]}$ mbar l s$^{-1}$ cm$^{-2}$ \cite{chiggiato}. This is the typical behaviour of metal surfaces discussed and interpreted by some authors\cite{fred,edwards,kanazawa}.
Quantitatively, both ferritic grades tested align with the values of AISI 304L, with the AISI 444 measuring the lowest outgassing rate after 10 h of pumping of 2.4$\times$10$^{-10}$ mbar l s$^{-1}$ cm$^{-2}$. The differences observed in the water vapour outgassing rate values could be due to the different roughness of the tested specimens rather than attributable to the grades.
\\
\begin{table}[h]
\caption{Water vapour specific outgassing rates measured at 21±2°C. Background removed.}
\begin{tabular}{ccc}
\hline
 Steel& \multicolumn{1}{c}{\begin{tabular}[c]{@{}c@{}}Tested samples\\ {[}cm$^{2}${]}\end{tabular}} & \multicolumn{1}{c}{\begin{tabular}[c]{@{}c@{}}q$_{10 h}$\\ {[}mbar l s$^{-1}$ cm$^{-2}${]}\end{tabular}}\\
\hline
AISI 441 & 4630 & 4.3$\times$10$^{-10}$\\
AISI 444 & 9260 & 2.4$\times$10$^{-10}$\\
AISI 304L & 5956 & 3.3$\times$10$^{-10}$\\
\hline
\label{tab:pumpdown}
\end{tabular}
\end{table}
\clearpage
\subsection{TEMPERATURE PROGRAMMED DESORPTION}
\begin{figure}[!h]
\includegraphics[scale=1]{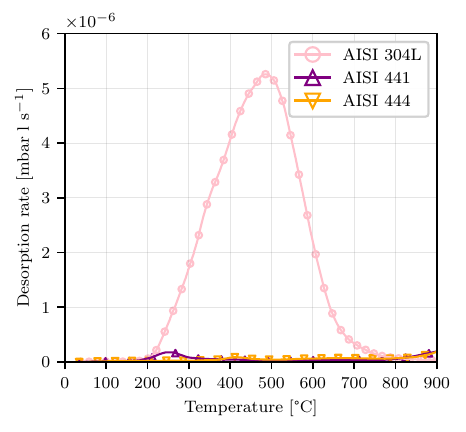}%
\caption{\label{fig:tpdtotal} H$_{2}$ thermal desorption spectra of AISI 304L, AISI 441 and AISI 444 samples. The background signal of the TPD system is removed. Heating rate: 5K/min.}%
\end{figure}

The H$_{2}$ thermal desorption mass spectra, with the background signal removed, are shown in \cref{fig:tpdtotal}. All raw data were smoothed for better visualization through a Savitzky-Golay filter\cite{filter} implemented in Python.\\
The signal obtained with AISI 304L samples can be fitted by a Fickian diffusion model that matches the broad peak with a maximum at 480°C (see \cref{fig:tpdtotal}). The obtained diffusion energy is 0.52±0.06 eV, i.e. a typical value in austenitic stainless steels\cite{TDS1,TDS2,TDS3}.\\ 
\begin{figure}[!h]
\includegraphics[scale=1]{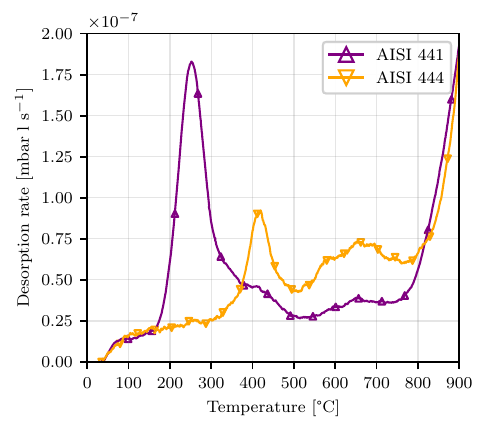}%
\caption{\label{fig:tpdferriticonly}H$_{2}$ thermal desorption spectra of AISI 441 and AISI 444 samples (vertical scale reduced x30 with respect to \cref{fig:tpdtotal}). The background signal of the TPD system is removed. Heating rate: 5K/min.}%
\end{figure}

The H$_{2}$ spectra of the two ferritic stainless steels show similar features (see \cref{fig:tpdferriticonly}). The AISI 441 samples are characterised by a narrow peak around 250-260°C, followed by a shoulder at 420°C. Additionally, the spectra feature a broad peak or shoulder between 530°C and 750°C before ramping up. Similarly, the AISI 444 samples exhibit the same peaks and shoulders, although the initial and highest peak is shifted to 420°C.

The shape of the peaks defining the ferritic stainless steel's H$_{2}$ TPD spectra appear not to follow Fickian desorption. They could be influenced by H de-trapping.

The high-intensity peak observed in the AISI 441 samples may be attributed to the production process of the raw materials. The residual dislocations from cold rolling in the AISI 441 samples could be accentuated by the subsequent skin passing required to achieve the 2B finish. In contrast, the AISI 444 samples, which were provided with a 2D finish and were not subjected to skin passing, did not exhibit this peak.

To further support the de-trapping limited H$_{2}$ desorption and the effect of the production processes, 0.3 cm thick hot-rolled AISI 441 samples were tested. The 0.3 cm samples had similar chemical composition and the same 2B surface finish as the 0.15 cm ones.

\begin{figure}[h]
\centering
\includegraphics[scale=1]{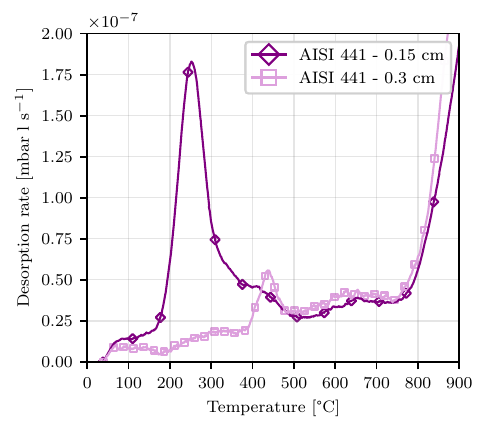}%
\caption{\label{fig:441thicknesses} H$_{2}$ thermal desorption spectra of AISI 441 0.15 cm and 0.3 cm thick samples. The former were skin passed while the latter were not. Heating rate: 5K/min.}%
\end{figure}

As depicted in \cref{fig:441thicknesses}, the 0.3 cm samples did not show the first main peak at around 250°C but rather a lower intensity at around 430°C, similar to what was observed for the AISI 444 samples. Therefore, the results seem to confirm the proposed interpretation of the data reported in \cref{fig:tpdferriticonly}.

It is important to note that if we exclude the main peak around 250°C for the 0.15 cm thick AISI 441 samples, which can easily be eliminated during a GWD bakeout operation, the position of the main peak appears to remain invariant regardless of the thickness and grade of the tested ferritic stainless steel.

The H$_{2}$ concentration in the measured samples is calculated by integrating the TPD signal up to 850°C and assuming uniform initial distribution in the volume of the samples. The selection of the integration range is arbitrary, justified by the observation that hydrogen released at temperatures above 850°C is tightly bound, rendering it non-diffusible within the temperature range relevant to most UHV applications. The calculation results are reported in \cref{tab:tpdferritic}. \\
\begin{table}[h]
\caption{AISI 444, AISI 441 and AISI 304L hydrogen concentration obtained by integrating the TPD spectra up to 850°C. The values reported are the average of at least three samples from the same batch. The background of the TPD system is removed. To convert atomic ppm to weight ppm, divide by 55.85 (molecular weight of iron). The AISI 304L samples are not vacuum fired.}
\begin{tabular}{ccc}
\hline
 Steel& \multicolumn{1}{c}{\begin{tabular}[c]{@{}c@{}}H content\\ {[}ppm at.{]}\end{tabular}} & \multicolumn{1}{c}{\begin{tabular}[c]{@{}c@{}}Thickness\\ {[}cm{]}\end{tabular}} \\
\hline
AISI 304L & 70 & 0.15\\
AISI 441 & 1.7 & 0.15\\
AISI 444 & 1.3 & 0.15\\
\hline
\label{tab:tpdferritic}
\end{tabular}
\end{table}

The calculated hydrogen concentration in AISI 304L samples is 41 to 53 times higher than that evaluated for the ferritic stainless steel samples, the latter showing H contents always below 2 atomic ppm. This outcome was expected due to the ferritic specimens' BCC structure, which has much lower hydrogen solubility than the austenitic FCC microstructure of AISI 304L.

\clearpage
\subsection{OUTGASSING RATES AFTER BAKEOUT}
The specific outgassing rates for H$_2$, CH$_4$, CO, and CO$_2$ are reported in \cref{tab:accu} and \cref{tab:accu1}. The values were calculated from data obtained by accumulation according to the procedure described in \cref{sec:accu}. The linearity of the accumulated gas quantity has been verified by measuring for different accumulation times.

As shown in \cref{tab:accu}, the selected ferritic grades showed exceptionally low outgassing rates after a 48 h bakeout at 80°C. The AISI 444 specimens had a H$_2$ specific outgassing rate of 1$\times$10$^{-15}$ mbar l s$^{-1}$ cm$^{-2}$, and for CH$_{4}$, CO and CO$_2$ values ranging between 1.8$\times$10$^{-16}$ and 6.1$\times$10$^{-16}$ mbar l s$^{-1}$ cm$^{-2}$.
In contrast, the AISI 441 samples showed a CH$_4$ specific outgassing rate of 1.1$\times$10$^{-16}$ mbar l s$^{-1}$ cm$^{-2}$, while the H$_2$, CO, and CO$_2$ outgassing rates fell below the detection limit, defined as 50\% of the background value. 

After the subsequent bakeout at 150°C for 48 hours (see \cref{tab:accu1}), the specific outgassing rates of AISI 441 fell below the system sensitivity, except for CH$_4$, which decreased by approximately one order of magnitude. For the AISI 444 samples, besides the H$_2$ outgassing rate being shadowed by the background value, CH$_4$, CO, and CO$_2$ showed reductions in outgassing by factors of 1.4 to 20 compared to the 80°C bakeout.

Compared to the similarly tested 304L samples, the selected ferritic alloys report H$_2$ specific outgassing rates from two to four orders of magnitude lower if considering the system sensitivity, therefore attaining values comparable if not lower than a mm thick austenitic stainless steel after vacuum firing (950°C, 2 h)\cite{chiggiato} or air bakeout (390°C, 100 h)\cite{bernardini}.

\begin{figure}[!h]
\includegraphics[scale=1]{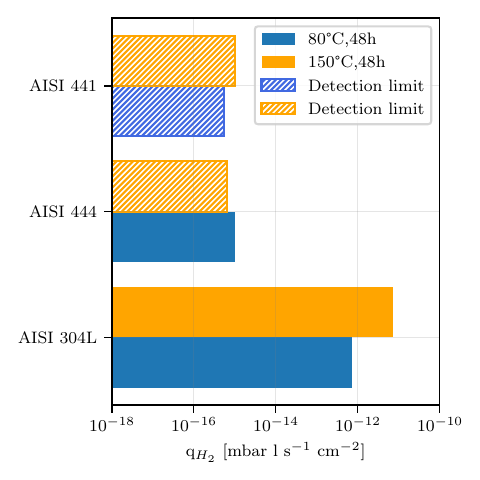}%
\caption{\label{fig:accuH}Comparison of H$_{2}$  specific outgassing rate reported in \cref{tab:accu,tab:accu1}. The system sensitivity (see definition in the text) normalised to the sample surface area is plotted as dashed columns when the measured values are below such a limit.}%
\end{figure}

Similarly to what was observed for mild steel \cite{mine} in a previous study, the ratios of hydrogen concentration and specific hydrogen outgassing rates for the ferritic grades, compared to austenitic steels, do not align quantitatively. As indicated by the TPD spectra discussed above, the hydrogen may be trapped within the surface oxide and the bulk of the material; thus, desorption is not limited by diffusion at room temperature.
It is worth noting the increase in the H$_2$ outgassing rate observed for AISI 304L when baked out at 80°C and 150°C. An analogous behaviour is evident in the background of these measurements (see \cref{fig:accuH}) and in line with observations reported by previous studies focused on vacuum fired austenitic stainless steels \cite{joustenfired}. However, the increase in the outgassing rate observed here for untreated 304L samples was significantly higher than what was previously reported for vacuum fired specimens. This trend, which might be relevant to the considered application, will be investigated in future work.\\

\begin{figure}[!h]
\includegraphics{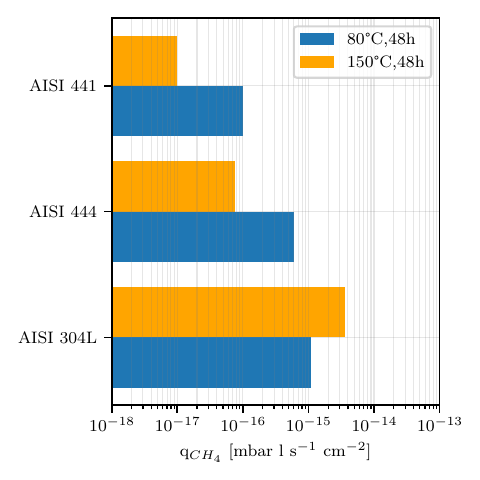}%
\caption{\label{fig:accuCH4}Comparison of CH$_{4}$  specific outgassing rate reported in \cref{tab:accu} and \cref{tab:accu1}. The system sensitivity (see definition in the text) normalised to the sample surface area is plotted as an dashed column when the measured values are below such a limit.}%
\end{figure}
\begin{figure}[!h]
\includegraphics[scale=1]{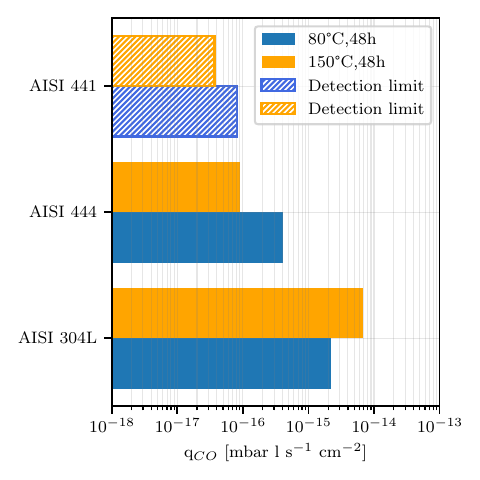}%
\caption{\label{fig:accuCO}Comparison of CO  specific outgassing rate reported in \cref{tab:accu} and \cref{tab:accu1}. The system sensitivity (see definition in the text) normalised to the sample surface area is plotted as an dashed column when the measured values are below such a limit.}%
\end{figure}
\begin{figure}[!h]
\includegraphics[scale=1]{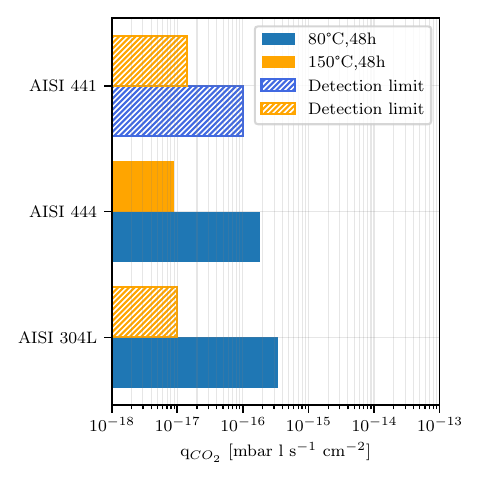}%
\caption{\label{fig:accuCO2}Comparison of CO$_{2}$  specific outgassing rate reported in \cref{tab:accu} and \cref{tab:accu1}.The system sensitivity (see definition in the text) normalised to the sample surface area is plotted as dashed columns when the measured values are below such a limit.}%
\end{figure}
\begin{table}[]
\caption{Specific outgassing rates for the selected stainless steels at 21±2°C after bakeout at 80°C for 48 h. Background signal removed. BSS: Below system sensitivity.}
\begin{tabular}{|c|cccc|}
\hline
\multirow{2}{*}{Steel grade} & \multicolumn{4}{c|}{\begin{tabular}[c]{@{}c@{}}Specific outgassing rate {[}mbar l s$^{-1}$ cm$^{-2}${]}\end{tabular}} \\ \cline{2-5} 
          & \multicolumn{1}{c|}{H$_{2}$} & \multicolumn{1}{c|}{CH$_{4}$} & \multicolumn{1}{c|}{CO} & CO$_{2}$ \\ \hline
AISI 441   & \multicolumn{1}{c|}{BSS} & \multicolumn{1}{c|}{1.1 $\times$ 10$^{-16}$}  & \multicolumn{1}{c|}{BSS}  & BSS   \\ \hline
AISI 444 & \multicolumn{1}{c|}{1.0 $\times$ 10$^{-15}$}  & \multicolumn{1}{c|}{6.1 $\times$ 10$^{-16}$}   & \multicolumn{1}{c|}{4.1 $\times$ 10$^{-16}$}   &  1.8 $\times$ 10$^{-16}$   \\ \hline
AISI 304L     & \multicolumn{1}{c|}{7.2 $\times$ 10$^{-13}$} & \multicolumn{1}{c|}{1.1 $\times$ 10$^{-15}$}  & \multicolumn{1}{c|}{2.2 $\times$ 10$^{-15}$}  & 3.4 $\times$ 10$^{-16}$   \\ \hline
\end{tabular}
\label{tab:accu}
\end{table}
\begin{table}[]
\caption{Specific outgassing rates for the selected stainless steels at 21±2°C after bakeout at 150°C for 48h. Background signal removed. BSS: Below system sensitivity.}
\begin{tabular}{|c|cccc|}
\hline
\multirow{2}{*}{Steel grade} & \multicolumn{4}{c|}{\begin{tabular}[c]{@{}c@{}}Specific outgassing rate {[}mbar l s$^{-1}$ cm$^{-2}${]}\end{tabular}} \\ \cline{2-5} 
          & \multicolumn{1}{c|}{H$_{2}$} & \multicolumn{1}{c|}{CH$_{4}$} & \multicolumn{1}{c|}{CO} & CO$_{2}$ \\ \hline
AISI 441   & \multicolumn{1}{c|}{BSS} & \multicolumn{1}{c|}{1.0 $\times$ 10$^{-17}$}  & \multicolumn{1}{c|}{BSS}  & BSS   \\ \hline
AISI 444 & \multicolumn{1}{c|}{BSS}  & \multicolumn{1}{c|}{4.3 $\times$ 10$^{-16}$}   & \multicolumn{1}{c|}{9.1 $\times$ 10$^{-17}$}   &  8.8 $\times$ 10$^{-18}$   \\ \hline
AISI 304L     & \multicolumn{1}{c|}{7.5 $\times$ 10$^{-12}$} & \multicolumn{1}{c|}{3.7 $\times$ 10$^{-15}$}  & \multicolumn{1}{c|}{6.7 $\times$ 10$^{-15}$}  & BSS   \\ \hline
\end{tabular}
\label{tab:accu1}
\end{table}
\clearpage
\section{CONCLUSIONS}
In this work, we explored the UHV compatibility of ferritic stainless steels and their applicability as structural materials for vacuum tubes of next-generation GWD.\\
For the selected samples, before bakeout, the water vapour outgassing rate follows the usual reciprocal function of the pumping time, showing values in line with those measured for austenitic stainless steels. Differences are considered to be related to different surface roughnesses more than to the grades of the steel.\\
The hydrogen content in ferritic stainless steel samples, analysed by TPD up to 850°C, is significantly lower—ranging from 41 to 53 times less—than the levels typically observed in as-cleaned AISI 304L. This disparity was anticipated due to the lower hydrogen solubility inherent in the body-centred cubic crystallographic structure characteristic of the alloys analysed. The TPD peaks indicate a more intricate desorption process than the typical behaviour seen in austenitic stainless steels, where hydrogen diffusion emerges as the dominant factor. The shape and positioning of these desorption peaks suggest that hydrogen liberation may be linked to de-trapping.\\
The bakeout temperatures chosen for this study are relatively low compared to the typical values used for UHV systems. This decision stems from the most practical and efficient method for baking the beampipe in GWD: utilizing the joule effect with electrical current applied to the vessel’s walls.
This study demonstrates that such low temperatures do not pose an issue for H$_{2}$ outgassing. Specifically, the measured H$_{2}$ specific outgassing rates, recorded at room temperature following a 48 h bakeout at 80°C, fall within the range of 10$^{-15}$ mbar l s$^{-1}$ cm$^{-2}$, comparable to those observed in vacuum-fired or air-baked austenitic stainless steel vacuum chambers. These values align with the requirements for future gravitational wave detectors and provide the benefit of avoiding expensive and time-consuming high-temperature degassing treatments.
While ferritic stainless steels offer an economical and high-performing alternative to austenitic stainless steels for UHV compatibility, positioning them as a promising baseline material for 3G GWDs vacuum pipes construction \cite{TDR}, their applicability from a manufacturing perspective is not yet fully established. 

Potential grain growth in the Heat Affected Zone (HAZ) of the weld line in ferritic stainless steels can lead to reduced mechanical strength in the welded area. This issue can be addressed by selecting stabilised ferritic stainless steel grades, such as AISI 441 or AISI 444, which contain titanium and niobium to limit grain growth during cooling. Additionally, careful optimization of welding parameters can help mitigate this effect. These considerations, along with a quantitative evaluation of corrosion resistance, are the focus of ongoing studies before ferritic stainless steels can be definitively established as the new baseline material for UHV beampipes in 3G GWDs.

\begin{acknowledgments}
We thank Mrs. Alice Michet and Mr. Jonathan Gaudio for their contribution to the outgassing measurements.\\
This work has been sponsored by the Wolfgang Gentner Programme of the German Federal Ministry of Education and Research (grant no. 13E18CHA)
\end{acknowledgments}

\section*{CONFLICT OF INTEREST}
The authors have no conflicts to disclose.

\section*{DATA AVAILABILITY}
The data that support the findings of this study are available from the corresponding author upon reasonable request.
\clearpage
\bibliography{mybib}

\end{document}